# The Non-Adiabatic Sub-Geometric Phase and It's Application on Quantum Transition

Zheng-Chuan Wang

School of Physical Sciences,

University of Chinese Academy of Sciences, Beijing 100049, China.

wangzc@ucas.ac.cn

## Abstract

Based on the adiabatic geometric phase concerning with density matrix[1] , we extend it to the sub-geometric phase in the non-adiabatic case. It is found that whatever the real part or imaginary part of the sub-geometric phase can play an important role in quantum transition. The imaginary part of sub-geometric phase can deviate the resonance peak in the quantum transition, which may bring modification on the level crossing, while the real part of sub-geometric phase will determine the stability of initial state according to the linear stability analysis theory, which can be regarded as somewhat complement on the selection rule of quantum transition. Finally, we illustrate them by two examples: one is the system with time-dependent perturbation, the other is a two-level system. It indicates that both the real and imaginary parts of sub-geometric phase have influence on quantum transition.

# I. Introduction

In 1984, Berry pointed out that a quantum system changing adiabatically with a slowing parameter will acquire a geometric phase in addition to the dynamical phase，when it cyclically return to it's initial state based on adiabatic theorem[2]. For the non-cyclic evolution with open paths in parameter space, Wu et al. generalized this phase and tested it by the muon spin resonance and level crossing resonance[3]. For the cyclic but non-adiabatic evolution, Aharonov and Anandan further extended the adiabatic Berry phase to the non-adiabatic case[4], they assume the wavefunction will acquire a total phase $\phi$ which contain both the non-adiabatic geometric phase and the dynamical phase after a period T. However, except some special states, i.e., the eigen-states or the Floquet state[5], the general wavefunction will not merely change a phase after a periodic evolution, because $|\psi(T)>$ and $|\psi(0)>$ may have different directions in Hilbert space, $|\psi(T)>\neq e^{i\phi}|\psi(0)>$. In 1988, the non-cyclic and non-unitary generalization of Aharonov-Anandan's non-adiabatic phase was given by Samuel and Bhandari[6], which was inspired by the Pancharatnam phase in the interference of polarized light[7]. However, the above adiabatic or non-adiabatic phases can be regarded as the phase accumulated during the evolution of state $|\psi(t)>$ after the dynamic phase is removed, so it can not appear in it's density matrix $\hat{\rho}=|\psi(t)><\psi(t)|$, because the geometric phase $e^{i\beta(t)}$ in $|\psi(t)>$ will cancel with the phase $e^{-i\beta(t)}$ in it's conjugate state $<\psi(t)|$. In 2013, Wang proposed a sub-geometric phase for the density matrix[1], it doesn't appear in the diagonal elements of density matrix, but appear in it's off-diagonal elements, which is just the difference of geometric phases between the two eigen-states of off-diagonal elements. However, this sub-geometric phase was given in the adiabatic and cyclic case, in this manuscript, we will extend it to the non-adiabatic and non-cyclic case.

As shown by Wu et al. in 1988[3], the geometric phase will have effect on the quantum transition, the resonant peak of muon spin resonance will deviate by an amount concerning with geometric phase. Usually, the quantum transition can be easily evaluated by time dependent perturbation theory when $H'(t)$ is very small than $H_0$ in the Hamiltonian $H = H_0 + H'(t)$. However, if $H'(t)$ is not a perturbation, usually it is very difficult to obtain an analytical solution for Schrödinger equation, and the quantum transition is not easy to analytically explore. For a simple linear time dependent Hamiltonian H=$A + Bt$, where $A$ and $B$ are $n \times n$ Hermite matrices, if $B$ is diagonal, it is refered to as the level crossing problem[8]. For the special case of $n = 2$, the transition probability can be analytically given by the Landau-Zener formula[9]. After adopting an independent crossing approximation, Brundobler and Elser extended the Landau-Zener formula to an arbitrary number of $n$ ($n \geq 2$)[8]. In this manuscript, we will explore the influence of sub-geometric phase on quantum transition.

As shown by Brundobler and Elser, the quantum transition between two states will occur when the corresponding energy levels are crossing[8]. Sometimes, the quantum transition will be forbidden according to the selection rule. In this manuscript, we will show that the sub-geometric phase, especially it's real part, is somewhat concerning with the selection rule. Usually, we are only interested in the imaginary part of geometric phase and don't care about it's real part, because it is not a real "phase". On the other hand, the geometric phase can be related with the quantum geometric tensor[10] whose imaginary part corresponds to the Berry curvature[11], and the real part corresponds to the quantum metric[12]. In this manuscript, we will study the influence of real part of sub-geometric phase on quantum transition by the theory of linear stability analysis[13], which is not only suit for investigating the classical non-equilibrim phase transition[14], but also exploited to study the stability of quantum system[15].

## II. Theoretical Formalism

Consider a quantum system with a Hamiltonian as $H = H_0 + H'(t)$, where the time independent Hamiltonian $H_0$ has it's eigen-state $\psi_k$ and eigen-value $E_k$ as

$$H_0\psi_k = E_k\psi_k, \qquad (1)$$

and $H'(t)$ is the time dependent term. Suppose the system remain in a eigen-state $\psi_k$ at initial time, after a time dependent disturbance $H'(t)$ is applied, the quantum transition will occur, it's wavefunction can then be written as

$$|\psi(t)> = \sum_n c_{nk}(t) e^{-iE_n t/\hbar} |\psi_n>, \qquad (2)$$

where the coefficient $c_{nk}(t)$ satisfies

$$i\hbar \dot{c}_{k'k}(t) = \sum_n e^{i\omega_{k'n} t} <k'|H'|n> \hbar c_{nk}, \qquad (3)$$

with $\omega_{k'n} = (E_{k'} - E_n)/\hbar$. Generally, $c_{nk}(t)$ is a complex function, we can express it as

$$c_{nk}(t) = A_{nk}(t) e^{i\varphi_{nk}(t)}, \qquad (4)$$

where $A_{nk}(t)$ is the amplitude of $c_{nk}(t)$ and $\varphi_{nk}(t)$ is it's phase. If we formally express $A_{nk}(t)$ as

$$A_{nk}(t) = e^{a_{nk}(t)}, \qquad (5)$$

where $a_{nk}(t) = \ln A_{nk}(t)$ is a real function, then Eq.(4) can be formally written as

$$c_{nk}(t) = e^{a_{nk}(t) + i\varphi_{nk}(t)}, \qquad (6)$$

the total "phase" $a_{nk}(t) + i\varphi_{nk}(t)$ is just our sub-geometric phase corresponding to the eigen-state $|\psi_n>$, which can be regarded as a generalization of Samuel and Bhandari's non-adiabatic and non-cyclic phase $\varphi_{S-B} = Im \int ds <\phi(s)|\frac{d}{ds}|\phi(s)>$[6] when we choose $\phi(s)$ as the eigen-state $c_{nk}(t) e^{-iE_n t/\hbar} |\psi_n>$ with the parameter $s$ as $t$, then

$$\varphi_{S-B} = (\varphi_{nk}(t) - E_n t/\hbar), \qquad (7)$$

after further eliminate the dynamic phase $-E_n t/\hbar$ as Samuel and Bhandari[6], $\varphi_{S-B} = \varphi_{nk}(t)$, so $\varphi_{nk}(t)$ is just the Samuel and Bhandari's phase corresponding to the eigen-state $|\psi_n\rangle$, but our sub-geometric phase has it's real part $a_{nk}(t)$ which is also useful. If we make a gauge transformation on the wavefunction

$$|\psi(t)\rangle \rightarrow e^{i\alpha}|\psi(t)\rangle = \sum_n e^{i\alpha} c_{nk}(t) e^{-iE_n t/\hbar}|\psi_n\rangle, \tag{8}$$

so

$$\varphi_{nk}(t) \rightarrow \varphi_{nk}(t) + \alpha, \tag{9}$$

$Im \langle \phi(t)|\frac{d}{dt}|\phi(t)\rangle$ satisfies the U(1) gauge transformation similar to Ref.[6], where $|\phi(t)\rangle = c_{nk}(t) e^{-iE_n t/\hbar}|\psi_n\rangle$. If we denote the set of states $\mathbb{N} = \{e^{i\alpha} c_{nk}(t) e^{-iE_n t/\hbar}|\psi_n\rangle\}$, $\mathbb{R}$ labels the space of rays: $\mathbb{R} = \mathbb{N}/\sim$, where $\sim$ represents the equivalent class which differ only by a phase in $\mathbb{N}$, then $(\mathbb{N}, \mathbb{R}, \pi)$ forms a principle fiber bundle over the base space $\mathbb{R}$, here π is a natural projection map π: $\mathbb{N} \rightarrow \mathbb{R}$. $Im \langle \phi(t)|\frac{d}{dt}|\phi(t)\rangle$ can be regarded as a connection of this U(1) principle fiber bundle, so our sub-geometric phase $\varphi_{nk}(t)$ has it's reasonable geometrical interpretation.

The difference between sub-geometric phase and Samuel and Bhandari's phase is: $\varphi_{S-B}$ is related to the whole wavefunction $|\psi(t)\rangle$, while the sub-geometric phase $\varphi_{nk}(t)$ is merely related to it's eigen-state $|\psi_n\rangle$, each eigen-state has it's own sub-geometric phase, there are many sub-geometric phases in the wavefunction $|\psi(t)\rangle$. The

Samuel and Bhandari's phase $e^{i\varphi_{S-B}}$ will not appear in the density matrix $\hat{\rho} = |\psi(t)><\psi(t)|$, because the phase $e^{i\varphi_{S-B}}$ in $|\psi(t)>$ and the phase $e^{-i\varphi_{S-B}}$ in $<\psi(t)|$ will cancel with each other, while the sub-geometric phase $\varphi_{nk}(t)$ can appear in the density matrix, it is

$$\hat{\rho} = \sum_{nn'} A_{nk} A_{n'k} e^{i(\varphi_{nk}(t)-\varphi_{n'k}(t))} |\psi_n><\psi_{n'}|, \qquad (10)$$

we can see that the sub-geometric phase will disappear in the diagonal elements, and exist in the off-diagonal elements as a difference $\varphi_{nk}(t) - \varphi_{n'k}(t)$ of sub-geometric phases between two eigen-states $|\psi_n>$ and $|\psi_{n'}>$. Substituting Eq.(10) into the Von-Neumann entropy $S = \text{Tr}\hat{\rho}ln\hat{\rho}$, the sub-geometric phases can also affect the Von-Neumann entropy. To the other physical observable, i.e., $<A> = \text{Tr}(\rho\hat{A})$, the sub-geometric phases in the density matrix will also have effect on the physical observable, which is a important character of our sub-geometric phase.

In the next, we will explore the application of sub-geometric phase on quantum transition. The imaginary part of sub-geometric phase can be combined together with the dynamic phase as $e^{i(\varphi_{nk}(t)-E_n t/\hbar)}$, because their evolution are all determined by the Schrödinger equation, which implies that the energy of system will be modified by the sub-geometric phase $\varphi_{nk}(t)$ as $(E_n - \frac{\hbar\varphi_{nk}(t)}{t})$. If we apply a periodical perturbation $H'(t)$ with frequency ω to the system, where ω=$(E_n - E_{k'})/\hbar$, then the resonant quantum transition will occur, but it's resonance peak will deviate from $(E_n - E_{k'})/\hbar$ to $(E_n - E_{k'} - \frac{\hbar\varphi_{nk}(t)}{t} + \frac{\hbar\varphi_{k'k}(t)}{t})/\hbar$, which is similar to the influence of adiabatic geometric phase on muon spin resonance[3]. The energy levels $(E_n - \frac{\hbar\varphi_{nk}(t)}{t})$ and $(E_{k'} - \frac{\hbar\varphi_{k'k}(t)}{t})$ are crossing during the quantum transition according to Brundobler and Elser's theory[8], and the crossing point will be modified by the corresponding sub-geometric phase.

In addition to the imaginary part of sub-geometric phase, the real part $a_{nk}(t)$ can also play an important role in quantum transition. If $H'(t)$ is a perturbation, the coefficient $c_{nk}(t)$ can be obtained by the time dependent perturbation theory: $c_{nk}(t) = \frac{1}{i\hbar}\int_0^t e^{i\omega_{nk}t'} H'_{nk} dt'$, $(n \neq k)$. In the special case $H'_{nk}=\langle\psi_n|H'|\psi_k\rangle=0$, the transition probability $P_{nk}(t) = |c_{nk}(t)|^2$ will be zero, which leads to the forbidden of quantum transition and selection rule. However, we will show that the quantum transition sometimes can not finally achieved even $H'_{nk} \neq 0$ and follow the selection rule, because it is affected by the real part $a_{nk}(t)$ of sub-geometric phase or the amplitude $A_{nk}(t)$ of $c_{nk}(t)$, too. According to the linear stability analysis theory[13], if $A_{nk}(t)=e^{a_{nk}(t)}$ increases with time $t$, then the initial state $|\psi_k\rangle$ will lose it's stability and the system will jump to the state $|\psi_n\rangle$. On the contrary, if $A_{nk}(t)$ decreases with time $t$, the initial state $|\psi_k\rangle$ will keep it's stability, the system will still stay in the initial state and doesn't jump to $|\psi_n\rangle$, in this case $H'_{nk} \neq 0$ and $c_{nk}(t) \neq 0$, but the quantum transition surely can not happen finally. So the real part $a_{nk}(t)$ can provide a complement to the selection rule. If $a_{nk}(t)$ is a constant and doesn't change with time $t$, it corresponds to the critical case between the stable and unstable. Finally, we conclude that the real part of sub-geometric phase determines the stability of initial state. Below, we will present two examples to expound the applications of sub-geometric phase.

## III. Examples:

### A. Time-dependent perturbation

If $H'(t)$ in the Hamiltonian is a perturbation $H'(t) = U(r,t)e^{i\omega t}$, i.e., for an amplitude tuned time dependent perturbation

$$H'(t) = U_\Omega(r) e^{i\Omega t} e^{i\omega t}, \tag{11}$$

where $\Omega$ is the frequency of amplitude tuned, which is much smaller than the oscillating frequency $\omega$, $\Omega \ll \omega$, $U_\Omega(r)$ is the amplitude. To the first order, the coefficient $c_{nk}(t)$ can be expressed by time dependent perturbation theory as

$$c_{nk}(t) = \frac{1}{i\hbar} \int_0^t e^{i(\omega_{nk}-\omega)t'} e^{i\Omega t'} <n|U_\Omega(r)|k> dt'. \tag{12}$$

Since $\Omega \ll \omega$, the term $e^{i\Omega t'}$ changes much slowly than the term $e^{i(\omega_{nk}-\omega)t'}$, so we can adopt Markov approximation $e^{i\Omega t'} \approx e^{i\Omega t}$, then Eq.(12) can be simplified as

$$c_{nk}(t) = <n|U_\Omega(r)|k> e^{i\Omega t} \times \frac{1}{i\hbar} \int_0^t e^{i(\omega_{nk}-\omega)t'} dt'$$

$$= \frac{<n|U_\Omega(r)|k>}{\hbar(\omega_{nk}-\omega)} e^{i\Omega t} (1 - e^{i(\omega_{nk}-\omega)t}). \tag{13}$$

If we write $c_{nk}(t) = A_{nk}(t) e^{i\varphi_{nk}(t)}$ as Eq.(4), the phase $e^{i(\omega_{nk}-\omega)t}$ in

Eq.(13) can be neglected when $\omega \approx \omega_{nk}$, it can be see that the time-dependent part of $\varphi_{nk}(t)$ is mainly contributed by $\Omega t$, and $A_{nk}(t)$ is mainly contribute by $Re(\frac{<n|U_\Omega(r)|k>}{\hbar(\omega_{nk}-\omega)})$ except a constant term. According to the linear stability analysis theory[13], since $Re(\frac{<n|U_\Omega(r)|k>}{\hbar(\omega_{nk}-\omega)})$ doesn't change with time t, the initial state $|\psi_k>$ possess the critical stability.

### B. Two-level system

If the time dependent term $H'(t)$ in Hamiltonian $H$ is not a perturbation, the solution for $c_{nk}(t)$ is usually difficult to find. Let us simply consider a two-level system with Hamiltonian $H = H_0 + H'(t)$, where $H_0 = \begin{pmatrix} -\Delta & 0 \\ 0 & \Delta \end{pmatrix}$ with eigen-energies $E_{0\pm} = \mp \Delta$ and eigen-states $|\phi_- >_0 = \begin{pmatrix} 1 \\ 0 \end{pmatrix}$ and $|\phi_+ >_0 = \begin{pmatrix} 0 \\ 1 \end{pmatrix}$, respectively, while $H'(t) = \begin{pmatrix} 0 & w(t) \\ w^*(t) & 0 \end{pmatrix} e^{-i\omega t}$ with $w(t) = |w(t)|e^{i\delta(t)}$, where $\delta(t) \ll \omega$, the wavefunction for the Hamiltonian $H$ can be written as

$$|\psi(t)> = \sum_{n=1,2} c_{nk}(t) e^{-iE_n t/\hbar}|\psi_n>. \tag{14}$$

Substituting it into the Schrödinger equation, we have

$$\dot{c}_{21}(t) = \frac{i}{\hbar}<2|H'|2>c_{21} - \frac{i}{\hbar}e^{i\omega_{21}t}<2|H'|1>c_{11}, \tag{15}$$

and

$$\dot{c}_{11}(t) = \frac{i}{\hbar}e^{i\omega_{12}t}<1|H'|2>c_{21} + <1|H'|1>c_{11}. \tag{16}$$

If we make a transformation on $c_{21}$ and $c_{11}$ as:

$$\tilde{c}_{21}(t) = c_{21}(t)\exp[\frac{i}{\hbar}\int_{-\infty}^{t} dt' <2|H'|2>], \tag{17}$$

and

$$\tilde{c}_{11}(t) = c_{11}(t)\exp[\frac{i}{\hbar}\int_{-\infty}^{t} dt'<1|H'|1>], \qquad (18)$$

then $\tilde{c}_{21}(t)$ and $\tilde{c}_{11}(t)$ satisfy

$$\dot{\tilde{c}}_{21}(t) = -\frac{i}{\hbar}e^{i\omega_{21}t}<2|H'|1>exp[\frac{i}{\hbar}\int_{-\infty}^{t}dt'(<2|H'|2>-<1|H'|1>]\tilde{c}_{11}(t), \qquad (19)$$

and

$$\dot{\tilde{c}}_{11}(t) = -\frac{i}{\hbar}e^{i\omega_{12}t}<1|H'|2>exp[\frac{i}{\hbar}\int_{-\infty}^{t}dt'(<1|H'|1>-<2|H'|2>]\tilde{c}_{21}(t). \qquad (20)$$

If we chose the initial condition as $c_{11}(t=-\infty)= 0$, from Eq.(20) we have

$$\tilde{c}_{11}(t) = \int_{-\infty}^{t} dt' [-\frac{i}{\hbar}e^{i\omega_{12}t'}<1|H'|2>exp[\frac{i}{\hbar}\int_{-\infty}^{t'}dt''(<1|H'|1>-<2|H'|2>)]\tilde{c}_{21}(t')], \qquad (21)$$

Substituting it into Eq.(19), we obtain

$$\dot{\tilde{c}}_{21}(t) = -\frac{i}{\hbar}e^{i\omega_{21}t}<2|H'|1>exp[\frac{i}{\hbar}\int_{-\infty}^{t}dt'(<2|H'|2>-<1|H'|1>)]$$
$$* \int_{-\infty}^{t} dt' [-\frac{i}{\hbar}e^{i\omega_{12}t'}<1|H'|2>exp[\frac{i}{\hbar}\int_{-\infty}^{t'}dt''(<1|H'|1>-<2|H'|2>)]\tilde{c}_{21}(t')]. \qquad (22)$$

If we adopt Markov approximation for $\tilde{c}_{21}(t')$ in the integral of the right hand side of Eq.(22) as[16]:

$$\tilde{c}_{21}(t') \approx \tilde{c}_{21}(t), \qquad (23)$$

we can express $\tilde{c}_{21}(t)$ under the initial condition $c_{21}(t=-\infty) = 1$ as:

$$\tilde{c}_{21}(t) = exp\{\int_{-\infty}^{t}\{-\frac{i}{\hbar}e^{i\omega_{21}t'}<2|H'|1>exp[\frac{i}{\hbar}\int_{-\infty}^{t'}dt''(<2|H'|2>-<1|H'|1>)]*\int_{-\infty}^{t'}dt''[-\frac{i}{\hbar}e^{i\omega_{12}t''}<1|H'|2>exp[\frac{i}{\hbar}\int_{-\infty}^{t''}dt'''(<1|H'|1>-<2|H'|2>)]]\}dt'\}.$$

(24)

Substituting the expression of H′(t) into Eq.(24) and Eq.(17), we have

$$c_{21}(t) = e^{a_{21}(t)+i\varphi_{21}(t)}, \quad (25)$$

where

$$a_{21}(t) = \frac{-1}{\hbar^2}\int_{-\infty}^{t}[|w(t')|\cos\delta(t')\int_{-\infty}^{t'}(|w(t'')|\cos(\delta(t'') - 2\omega_{21}t''))dt'' + |w(t')|\sin\delta(t')\int_{-\infty}^{t'}(|w(t'')|\sin(\delta(t'') - 2\omega_{21}t''))dt'']dt'$$

(26)

and

$$\varphi_{21}(t) = \frac{-1}{\hbar^2}\int_{-\infty}^{t}[|w(t')|\cos\delta(t')\int_{-\infty}^{t'}(|w(t'')|\sin(\delta(t'') - 2\omega_{21}t''))dt'' - |w(t')|\sin\delta(t')\int_{-\infty}^{t'}(|w(t'')|\cos(\delta(t'') - 2\omega_{21}t''))dt'']dt',$$

(27)

where $a_{21}(t)$ and $\varphi_{21}(t)$ are the real and imaginary parts of the sub-geometric phase, respectively. The real part will have influence on the transition probability

$$P_{21}(t) = |c_{21}(t)|^2 = e^{2a_{21}(t)}, \quad (28)$$

and the imaginary part will deviate the resonance peak from $\omega = (E_2\text{-}E_1)/\hbar$ to $(E_2 - E_1 - \frac{\hbar\varphi_{21}(t)}{t} + \frac{\hbar\varphi_{11}(t)}{t})/\hbar$.

We can further discuss the special case $w(t) = B_0 e^{\lambda t}$, where $B_0$ and

$\lambda$ are positive constant. Since $\delta(t)$ varies much slowly than $\omega t$, one can approximate it as a constant $\delta(t) \approx \delta_0$, for simplicity we only set $\delta_0=0$, then

$$a_{21}(t)=-\frac{1}{\hbar^2}\frac{B_0^2}{(\lambda^2+4\omega_{21}^2)}[\frac{\lambda e^{2\lambda t}}{4\lambda^2+4\omega_{21}^2}(2\lambda\cos(2\omega_{21}t)+2\omega_{21}\sin(2\omega_{21}t))$$

$$+\frac{2\omega_{21}e^{2\lambda t}}{4\lambda^2+4\omega_{21}^2}(2\lambda\sin(2\omega_{21}t)-2\omega_{21}\cos(2\omega_{21}t))], \qquad (29)$$

and

$$\varphi_{21}(t) = \frac{1}{\hbar^2}\frac{B_0^2}{(\lambda^2+4\omega_{21}^2)}[\frac{(2\lambda^2-4\omega_{21}^2)e^{2\lambda t}}{4\lambda^2+4\omega_{21}^2}\sin(2\omega_{21}t) - \frac{6\omega_{21}\lambda e^{2\lambda t}}{4\lambda^2+4\omega_{21}^2}\cos(2\omega_{21}t)].$$

$$(30)$$

Since $e^{a_{21}(t)}$ decreases with time $t$, then the state $|1>$ is stable according to the linear stability theory, because we adopt the initial condition as $c_{11}(t=-\infty)= 0$ and $c_{21}(t =-\infty) = 1$, the system start it's evolution from the state $|2>$ at t $=-\infty$, under the disturbance $H'(t)$ it will arrive at the state $|1>$ and finally stay here, so the state $|1>$ is stable.

### IV. Summary and Discussions

In this paper, we propose a non-adiabatic and non-cyclic sub-geometric phase in the time dependent system, it can be regarded as a generalization of Samuel and Bhandari's phase[6] which has only one phase corresponding to a quantum state, while there are many sub-geometric phases corresponding to each eigen-state in our formalism. On the other hand, Samuel and Bhandari's phase only concentrate on the imaginary part of wavefunction during the time evolution, our sub-geometric phase is interested in it's real part, too. We show that the real part of sub-geometric phase has influence on the stability of the initial state according to the linear stability analysis theory. Sometimes it will decay the quantum transition, and can be considered as a complement to the selection rule. While the imaginary part of sub-geometric phase will have a modification on the resonance peak of quantum transition which corresponds to the level crossing.

We also further illustrate the sub-geometric phase by two examples: when the $H'(t)$ in the Hamiltonian is a perturbation, we discuss the sub-geometric phase by time dependent perturbation theory; when $H'(t)$ is not a perturbation, we simply discuss a two-level system. Both of them demonstrate clearly the influence of sub-geometric phase on quantum transition. It should be pointed out that although the Landau-Zener problem can be extended to the system with many levels ($n \geq 2$) when $H'(t)$ is not a perturbation[8], but how to extend our sub-geometric phase from the above two-level system to the many levels case is still an open question, because it's Hamiltonian is different from the Landau-Zener's, we leave it for further exploration.

**Acknowledgments**

This study is supported by the National Key R&D Program of China (Grant No. 2022YFA1402703), the Strategic Priority Research Program

of the Chinese Academy of Sciences (Grant No. XDB28000000).## References

[1] Z. C. Wang, Scientific Reports, 9, 13258 (2019).

[2] M. V. Berry, Proc. R. Soc. London A 392, 45 (1984).

[3] Y. S. Wu and H. Z. Li, Phys. Rev. B 38, 11907 (1988).

[4] Y. Aharonov and J. Anandan, Phys. Rev. Lett. 58, 1593 (1987).

[5] J. H. Shirley, Phys. Rev. 138, B979 (1965).

[6] J. Samuel and R. Bhandari, Phys. Rev. Lett. 60, 2339 (1988).

[7] S. Pancharatnam, Proc. Indian Acad. Sci. A44, 247 (1956).

[8] S. Brundobler and V. Elser, J. Phys. A: Math. Gen. 26, 1211 (1993).

[9] C. Zener, Proc. R. Soc. A137, 696 (1932).

[10] W. Chen and G. von Gersdorff, SciPost Phys. Core 5, 040 (2022).

[11] M. V. Berry, Geometric Phases in Physics, World Scientific, Singapore, (1989).

[12] J. P. Provost and G. Vallee, Comm. Math. Phys.76, 289 (1980).

[13] R. Bellman, Stability theory of Differential equations, McGraw-Hill, New York 1953.

[14] D. H. Sattinger, Topics in stability and Bifurcation theory, Springer, Berlin, 1973.

[15] Z. C. Wang, Mod. Phys. Lett. B36, 2250037 (2022).